\documentclass[10pt,conference]{IEEEtran} 
\IEEEoverridecommandlockouts

\usepackage{subfiles} 
\usepackage{multirow}
\usepackage[ruled,linesnumbered]{algorithm2e}
\usepackage{listings}
\usepackage{tcolorbox}
\usepackage{subcaption}

\usepackage{subfloat}
\usepackage{color,xcolor,colortbl}
\usepackage{xspace}
\usepackage{enumitem}
\usepackage{graphicx}
\usepackage{xr}
\usepackage{url}
\usepackage{booktabs}
\usepackage{diagbox}
\usepackage{array}
\usepackage{cite}
\usepackage{amssymb}
\usepackage{tcolorbox}

\makeatletter
\newcommand{\thickhline}{%
    \noalign {\ifnum 0=`}\fi \hrule height 1pt
    \futurelet \reserved@a \@xhline
}
\newcolumntype{"}{@{\hskip\tabcolsep\vrule width 1pt\hskip\tabcolsep}}
\makeatother

\newcommand{\namei}{TGS\xspace}
\newcommand{\nameii}{VGS\xspace}
\newcommand{\final}{GS\xspace}
\newcommand{\http}{\url{https://github.com/AMPLE001/AMPLE}}

\newcommand{\ksr}{kernel-scaled representation module\xspace}
\newcommand{\ksrab}{KSR\xspace}

\newcommand{\tool}{AMPLE\xspace}
\newcommand{\eg}{\textit{e}.\textit{g}.\xspace}
\newcommand{\ie}{\textit{i}.\textit{e}.\xspace}
\newcommand{\et}{\textit{et} \textit{al}\xspace}
\newcommand{\ivd}{IVDetect\xspace}

\AtBeginDocument{%
  \providecommand\BibTeX{{%
    \normalfont B\kern-0.5em{\scshape i\kern-0.25em b}\kern-0.8em\TeX}}}

\begin{document}
\title{Vulnerability Detection with Graph Simplification and Enhanced Graph Representation Learning}
\author{\IEEEauthorblockN{Xin-Cheng Wen$^{1\star}$, Yupan Chen$^{1\star}$, Cuiyun Gao$^{1\ast}$, Hongyu Zhang$^{2}$, Jie M. Zhang$^{3}$, Qing Liao$^{1}\circ$}

\IEEEauthorblockA{$^1$ School of Computer Science and Technology, Harbin Institute of Technology, Shenzhen, China}

\IEEEauthorblockA{$^2$ School of Big Data and Software Engineering, Chongqing University, China}

\IEEEauthorblockA{$^3$ Department of Informatics, King's College London, UK}

\IEEEauthorblockA{xiamenwxc@foxmail.com,
cyp36889@gmail.com,
gaocuiyun@hit.edu.cn, \\
hyzhang@cqu.edu.cn, 
jie.zhang@kcl.ac.uk,
liaoqing@hit.edu.cn
}

\thanks{$^\star$ These authors contribute to the work equally and are co-first authors of the paper.} 

\thanks{$^{\ast}$ Corresponding author. The author is also affiliated with Peng Cheng Laboratory and Guangdong Provincial Key Laboratory of Novel Security Intelligence Technologies.}
\thanks{$^{\circ}$ The author is also affiliated with Peng Cheng Laboratory.}
}
\pagestyle{plain}
\maketitle

\begin{abstract}
Prior studies have demonstrated the effectiveness of Deep Learning (DL) 
in automated software vulnerability detection. Graph Neural Networks (GNNs) have proven effective in learning the graph representations of source code and are commonly adopted by existing DL-based vulnerability detection methods. However, the existing methods are still limited by the fact that GNNs are essentially difficult to handle the connections between long-distance nodes in a code structure graph. Besides, they do not well exploit the multiple types of edges in a code structure graph (such as edges representing data flow and control flow). Consequently, despite achieving state-of-the-art performance, the existing GNN-based methods tend to fail to capture global information (\ie, long-range dependencies among nodes) of code graphs. 

To mitigate these issues, in this paper, we propose a novel vulnerability detection framework with gr\textbf{A}ph si\textbf{M}plification and enhanced graph re\textbf{P}resentation \textbf{LE}arning, named \textbf{\tool}. \tool mainly contains two parts: 1) graph simplification, which aims at reducing the distances between nodes by shrinking the node sizes of code structure graphs; 2) enhanced graph representation learning, which involves one edge-aware graph convolutional network module for fusing heterogeneous edge information into node representations and one \ksr for well capturing the relations between distant graph nodes. Experiments on three public benchmark datasets show that \tool outperforms the state-of-the-art methods by 0.39\%-35.32\% and 7.64\%-199.81\% with respect to the accuracy and F1 {score} metrics, respectively. The results demonstrate the effectiveness of \tool in learning global information of code graphs for vulnerability detection.

\end{abstract}


\section{Introduction}

Software vulnerabilities are weaknesses in source code that can be exploited to cause safety issues such as user information leakage~\cite{Exactis} and cyber extortion~\cite{ransomware_attack}. The number of disclosed vulnerabilities constantly increases, causing more and more significant concerns in the field of software industry and cybersecurity.
For example, the National Vulnerability Database (NVD) in the US~\cite{nvd} released 8,051 vulnerabilities in the first quarter of 2022, with an increase of 25\% over the same period last year~\cite{statistics}. 
Most recently, Synopsys Open Source Security and Risk Analysis (OSSRA) analyzed 2,409 codebases and found that 81\% of them contained at least one known open source vulnerability~\cite{OSSRA}. The severity and prevalence of software vulnerabilities make it critical to develop
accurate automatic vulnerability detection techniques.

Conventional vulnerability detection methods~\cite{s1, s2, s3, s4, s5} mainly adopt pre-defined rules provided by human experts to conduct code analysis, which are labor intensive and inaccurate. Recently, many Deep Learning (DL)-based vulnerability detection methods have been proposed~\cite{vuldeepecker, russell, sysevr, dam2017automatic, vulcnn}, which achieve state-of-the-art performance in vulnerability detection via automatically learning the patterns of vulnerable code without human heuristics. The DL-based methods~\cite{vuldeepecker, russell, sysevr, dam2017automatic} generally leverage the structural information of source code and represent code as a \textit{code structure graph}
such as Control Flow Graph (CFG), Data Flow Graph (DFG), and Code Property Graph (CPG)~\cite{cpg}. The information in these graphs can also be integrated 
with Abstract Syntax Tree (AST) and Natural Code Sequence (NCS) to provide more comprehensive graph representations~\cite{devign}.

Code structure graphs typically contain complex hierarchical information~\cite{graphcodebert,DBLP:conf/iclr/NguyenJHS20}. To effectively learn the representations of code structure graphs, state-of-the-art DL-based methods utilize a variety of Graph Neural Network (GNN) models such as Gated Graph Neural Network (GGNN)~\cite{reveal} and Graph Convolution Network (GCN)~\cite{IVDETECT}. However, it is well acknowledged that GNN models have limitations in handling long-distance connections between nodes that are not directly adjacent to each other~\cite{zhu2021pre,hellendoorn2019global}. Although one can use GNNs with stacked multiple layers to learn global information (\ie, long-range dependencies among nodes) of the graph, the over-smoothing issue~\cite{bottleneck,li2018deeper} yielded from a deep GNN
will lead to similar embeddings for nodes with different labels, thereby degrading
the overall performance~\cite{geom-gcn}.
Moreover, the code structure graphs typically contain multiple types of relations (\eg, edges representing data flow and control flow) \cite{reveal,IVDETECT}. The existing GNN models (\eg GGNN, GCN~\cite{GCN-1}) cannot adequately model the edge features in a graph~\cite{edge_drawback}, which further limits their capability to capture global information of a code structure graph. Therefore, despite achieving state-of-the-art performance, existing DL-based models still have difficulties in capturing the global information of source code, which hinders their performance in detecting software vulnerabilities especially for complex graphs.

\begin{figure}[t]
\centering
\includegraphics[width=0.50\textwidth]{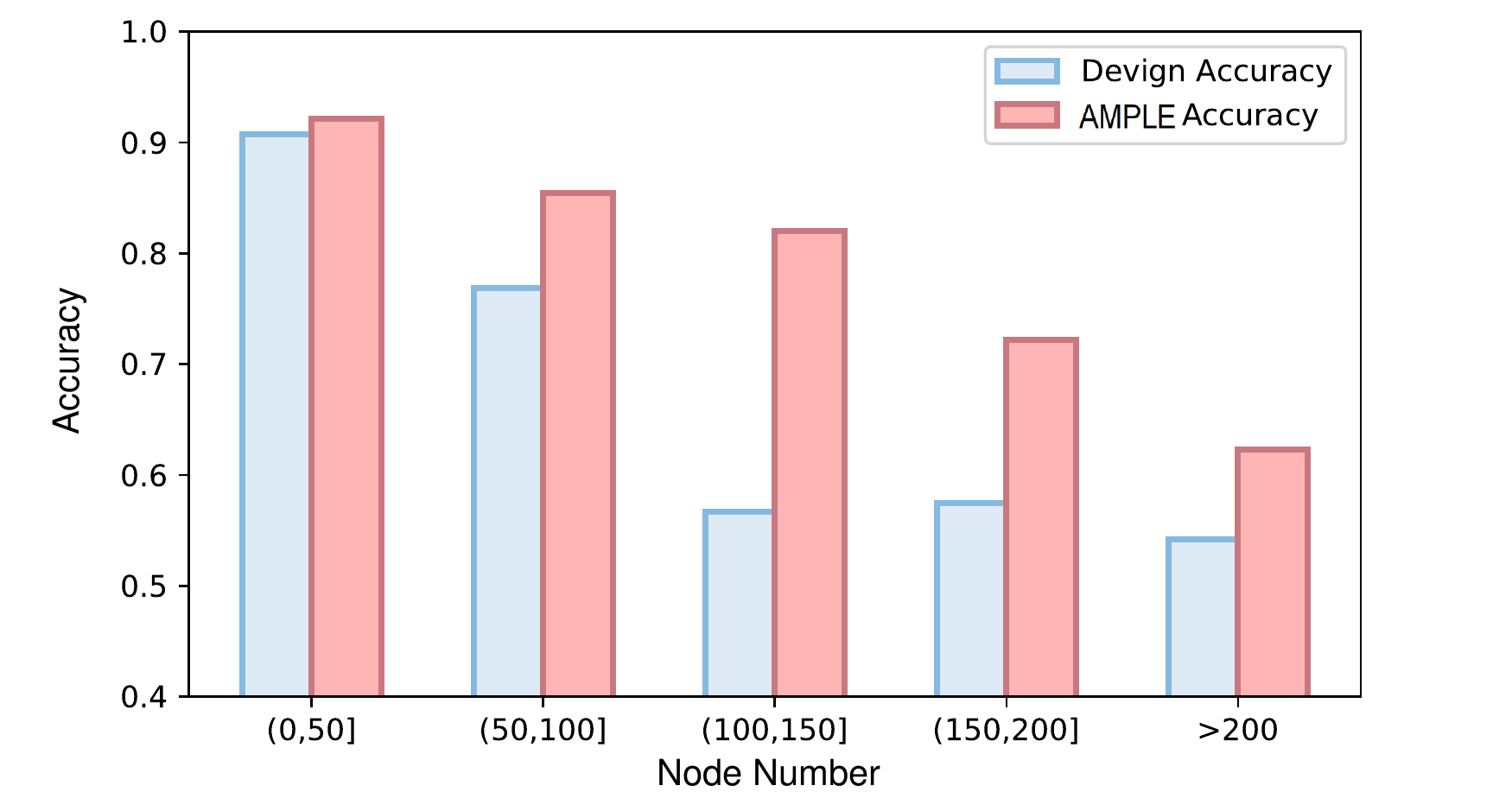}
\caption{Accuracy of Devign~\cite{devign} and \tool on the Reveal dataset~\cite{reveal} with different numbers of graph nodes. }
\label{discussion}
\end{figure}

Taking Devign~\cite{devign}, a recent GNN-based vulnerability detection model as an example, we analyze the relationship between model performance and the number of graph nodes.
Devign learns vulnerable code patterns from code structure graphs that contain four types of edges (AST, CFG, DFG, and NCS). We experiment on the Reveal dataset \cite{reveal}. The data in the test set are divided into five intervals based on the number of nodes in code structure graphs, with results shown in Figure~\ref{discussion}. As can be seen, Devign shows obviously better performance for the graphs with fewer nodes, \eg, achieving 90\% accuracy for graphs with no more than 50 nodes, and drops severely as the number of nodes increases. For the graphs with more than 200 nodes, the accuracy of Devign is only around 54\%. The results indicate the ineffectiveness of current methods in learning global code information. 




In this paper, we propose \textbf{\tool}, a novel vulnerability detection framework with gr\textbf{A}ph si\textbf{M}plification and enhanced graph re\textbf{P}resentation \textbf{LE}arning. \tool contains two major parts: 1) \textbf{Graph simplification}, which reduces the distances between nodes by shrinking the sizes of code structure graphs.
2) \textbf{Enhance graph representation learning}, which includes two modules. The edge-aware graph neural network module considers the information of edge types and fuses the heterogeneous edge information into node representations; while the kernel-scaled representation module scales up the convolution kernel size for explicitly capturing the relationship between distant graph nodes.

To evaluate \tool{}, we use three widely-studied benchmark datasets in software vulnerability detection: {FFMPeg+Qemu~\cite{devign}, Reveal\cite{reveal}, and Fan et al.~\cite{fan}}. We compare \tool{} with six existing software vulnerability detection methods, including three state-of-the-art graph-based 
and three token-based methods. 
The results demonstrate that \tool outperforms all the baseline methods with respect to the accuracy and F1 score metrics.
In particular, \tool{} achieves 4.75\%, 6.86\%, and 9.24\% absolute improvement in F1 score over the state-of-the-art
on the three datasets, respectively, with the corresponding relative improvements at 7.63\%, 16.48\%, and 40.40\%.

In summary, the major contributions of this paper are as follows:
\begin{enumerate}


\item We propose \tool, a novel vulnerability detection framework. \tool involves a new code input representation for reducing the distances between nodes and an enhanced graph representation learning method for better capturing the information code structure graphs.

\item We perform a large-scale evaluation of \tool on three public benchmark datasets, and the results demonstrate the effectiveness of \tool in accurate vulnerability detection.
\end{enumerate}

The remaining of this paper is organized as follows. Section~\ref{sec:background} introduces the background and related work. Section~\ref{sec:framework} presents the overall architecture of \tool and the two parts in detail, including graph simplification and enhanced graph representation learning. Section~\ref{sec:experimentalsetup} describes the experimental setup, including datasets, baselines and experimental settings. Section~\ref{sec:results} introduces the experimental results and analysis. Section~\ref{sec:discussion} discusses why \tool can effectively detect code vulnerability and the threats to validity.  Section~\ref{sec:conclusion} concludes this paper.

\section{Background and Related Work}
\label{sec:background}
\begin{figure*}[ht]
	\centering
	\includegraphics[width=1.\textwidth]{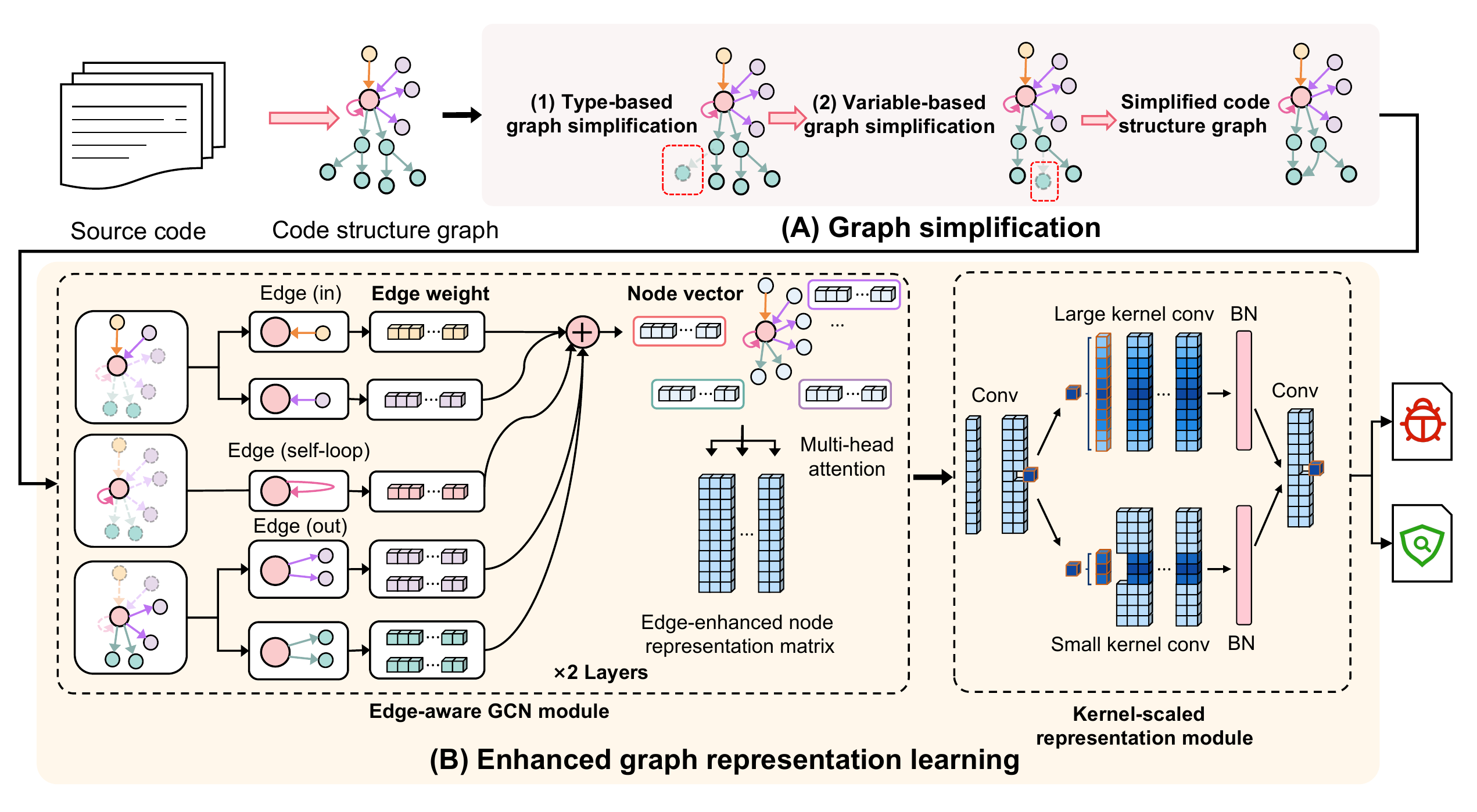}
	\caption{
	Architecture of \tool, which mainly contains two parts: (A) graph simplification and (B) enhanced graph representation learning. Nodes with different colors in the simplified code structure graph indicate their edge types are different, \eg, edges of CFG or DFG. 
	Conv \cite{CNN1} represents convolutional layer. 
	BN \cite{ioffe2015batch} is the batch normalization layer.
	}
	
	\label{architecture}
\end{figure*}

In this section, we introduce the background and related work about vulnerability detection, from the perspectives of the code representation and deep learning models.

\subsection{Code Representation in Vulnerability Detection}
Early DL-based vulnerability detection methods~\cite{vuldeepecker,russell, dam2017automatic,DBLP:journals/corr/abs-2207-11104,DBLP:conf/sigsoft/WangYGP0L22} regard source code as flat sequences and adopt natural language processing techniques for representing the input code. They generally initialize the embeddings of tokens with word2vec~\cite{wv}. Despite that these methods do not require code integrity and syntax correctness, they neglect the inherent structural regularity of code, and 
may only focus on shallow semantics~\cite{devign}. To well exploit the structural information of code, a series of methods~\cite{allamanis2017learning, sysevr, IVDETECT, reveal, duan2019vulsniper, vulcnn, hgvul, DBLP:journals/corr/abs-2104-09340, DBLP:conf/msr/MaZSHZPCXT22} abstract the code in the form of graphs (\eg AST, CFG, DFG, PDG) and use the graphs as the model input. {Following the work of Devign~\cite{devign}, we 
construct} a comprehensive code structure graph, which is treated as a heterogeneous graph to obtain the semantic and syntactic information represented by different edge types. {We further simplify the code structure graph to shrink the graph size and boost the model to learn the global representation.}

\subsection{Deep-learning Models for Vulnerability Detection}

The neural networks used in early work include Convolutional Neural Network (CNN) ~\cite{DBLP:conf/nips/KrizhevskySH12, DBLP:journals/pieee/LeCunBBH98} and Recurrent Neural Network (RNN) \cite{DBLP:journals/air/AlhumoudW22, DBLP:conf/slt/MikolovZ12}. 
For example, Russell \et.~\cite{russell} input the lexical code representation into CNN. Vuldeepecker \cite{vuldeepecker} employs BiLSTM to handle the code gadgets, a fine-grained code slice. Some work \cite{gong2019joint,sysevr} also selects LSTM as code encoders.
Recently, GNN-based methods~\cite{devign, reveal,IVDETECT} have achieved state-of-the-art performance on vulnerability detection. {GNN~\cite{GNN-2,DBLP:journals/tnn/ScarselliGTHM09} is a type of neural network which directly operates on graph structure.} 
Devign \cite{devign} and Reveal \cite{reveal}
adopt GGNN \cite{Method3} to process multiple directed graphs generated from source code. {However, GGNN converts different edge types into an adjacency matrix without distinguishing them, leading to limited performance as the number of edge types increases~\cite{cao2022mvd}.}
\ivd~\cite{IVDETECT} leverages Graph Convolutional Network (GCN) and  uses a context vector to predict vulnerabilities. However, GCN {focuses on learning}
the local features of nodes by neighborhood aggregation{, and thereby tends to fail to capture long-range dependencies among nodes as the graph sizes increase~\cite{fu2022linevul}.}
In this work, we {aim at mitigating the issues of GNNs for better learning the global information of code structure graphs in vulnerability detection.}

\section{\tool Framework}
\label{sec:framework}

In this section, we introduce the detailed architecture of \tool. As shown in Figure~\ref{architecture},
\tool mainly consists of two parts: (A) Graph simplification ($c.f.$, Section \ref{GraphSimplification}), which condenses the {repetitive information in} code structure graph through type-based graph simplification and variable-based graph simplification; (B) Enhanced graph representation learning, which contains the edge-aware graph convolutional network
module ($c.f.$, Section \ref{Edge-aware}) to learn edge-enhanced node representations and the kernel-scaled representation
module ($c.f.$, Section \ref{Kernel-scaled}) to capture the relations between distant graph nodes.

\begin{table}[ht]
\centering

\setlength{\tabcolsep}{1.2mm}
\renewcommand{\arraystretch}{1.1}
\caption{{Merging rules in type-based graph simplification.} ``PType'' and ``CType'' indicate the types of the parent and child nodes of a directed edge, respectively. The {*} represents any node type, and the PType ``Augment'' in Rule 6 represents the parameters of function call statements.}
\scalebox{0.95}{\begin{tabular}{|c|c|c|}
\hline
Rule 	& PType	 & CType    \\
\hline
\multirow{4}{*}{1} & \multirow{4}{*}{ExpressionStatement} & AssignmentExpression          \\
                   &                                      & UnaryExpression               \\
                   &                                      & CallExpression                \\
                   &                                      & PostIncDecOperationExpression \\
\hline
2		 &IdentifierDeclStatement & IdentifierDecl \\
\hline
3        & Condition      & *   \\
\hline
4       & ForInit   &     *  \\
\hline
5        & CallExpression    &  ArgumentList    \\
\hline
6        & Argument  &  *     \\
\hline
7        &Callee    &Identifier     \\
\hline

\end{tabular}
}

\label{type}
\end{table}



\subsection{Graph Simplification} \label{GraphSimplification}




\begin{figure}[t]
\centering
\includegraphics[width=0.44\textwidth]{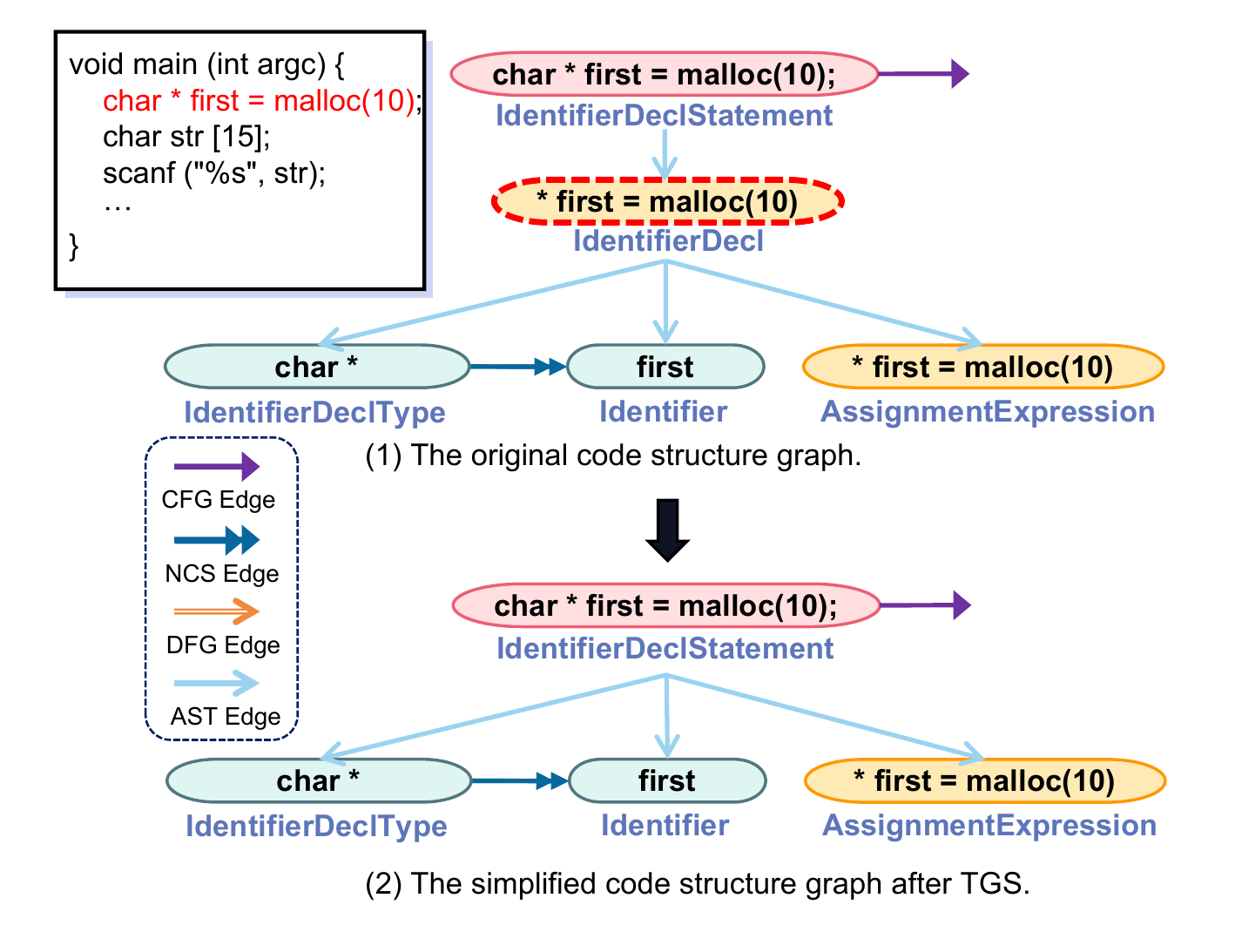}
\caption{An example illustrating  type-based graph simplification. 
{The text in/under each node represents the node value/type.} Some nodes and edges are omitted {for facilitating illustration}. The node with red dashed border in {the original code structure graph (1)}
is merged with its parent node, producing a TGS-based simplified graph (2).}
\label{cs_1}
\end{figure}


The graph simplification part aims at condensing the repetitive information in a code structure graph, which can shrink graph sizes and reduce the distances between nodes. 
We propose two simplification steps: 
type-based graph simplification (according to node types) followed by variable-based graph simplification (according to node variables).
In the following, we first elaborate on the two steps, and then provide an algorithm of the overall process.

\subsubsection{Type-based Graph Simplification}
The type-based graph simplification (TGS) method aims at merging adjacent nodes according to the node types. According to the parsing principles~\cite{yamaguchi2014modeling} and manually examining the code structure graphs,  
we have identified seven merging rules for type-based graph simplification.
{Table~\ref{type} lists the summarized merged rules, where} \textit{PType} and \textit{CType} denote the type of the parent node and child node, respectively. The merging rule 1, 2, 3, 4 and 5-7 {correspond to}
different {types of} statements {including} expression statement, identifier declaration statement, condition statement, for-loop statement and function call statement, respectively, in {the studied C/C++ programming language}.
For each pair of adjacent nodes matching one merging rule, the child node will be removed since its information is the refinement of its parent node
and can also be reflected in its subsequent nodes. 
Figure~\ref{cs_1} illustrates an example of the TGS-based simplification process. In Figure~\ref{cs_1} (1),
the node with red dashed border is merged to its parent node according to Rule 2 in Table~\ref{type}. Specifically, the information in the child node ``\textit{* first = malloc(10)}'' is also covered by the parent node and its three child nodes. Due to the page limit, we illustrate the simplification based on other rules in the published repository\footnote{{\http}}.

\begin{figure}[t]
\centering

\includegraphics[width=0.5\textwidth]{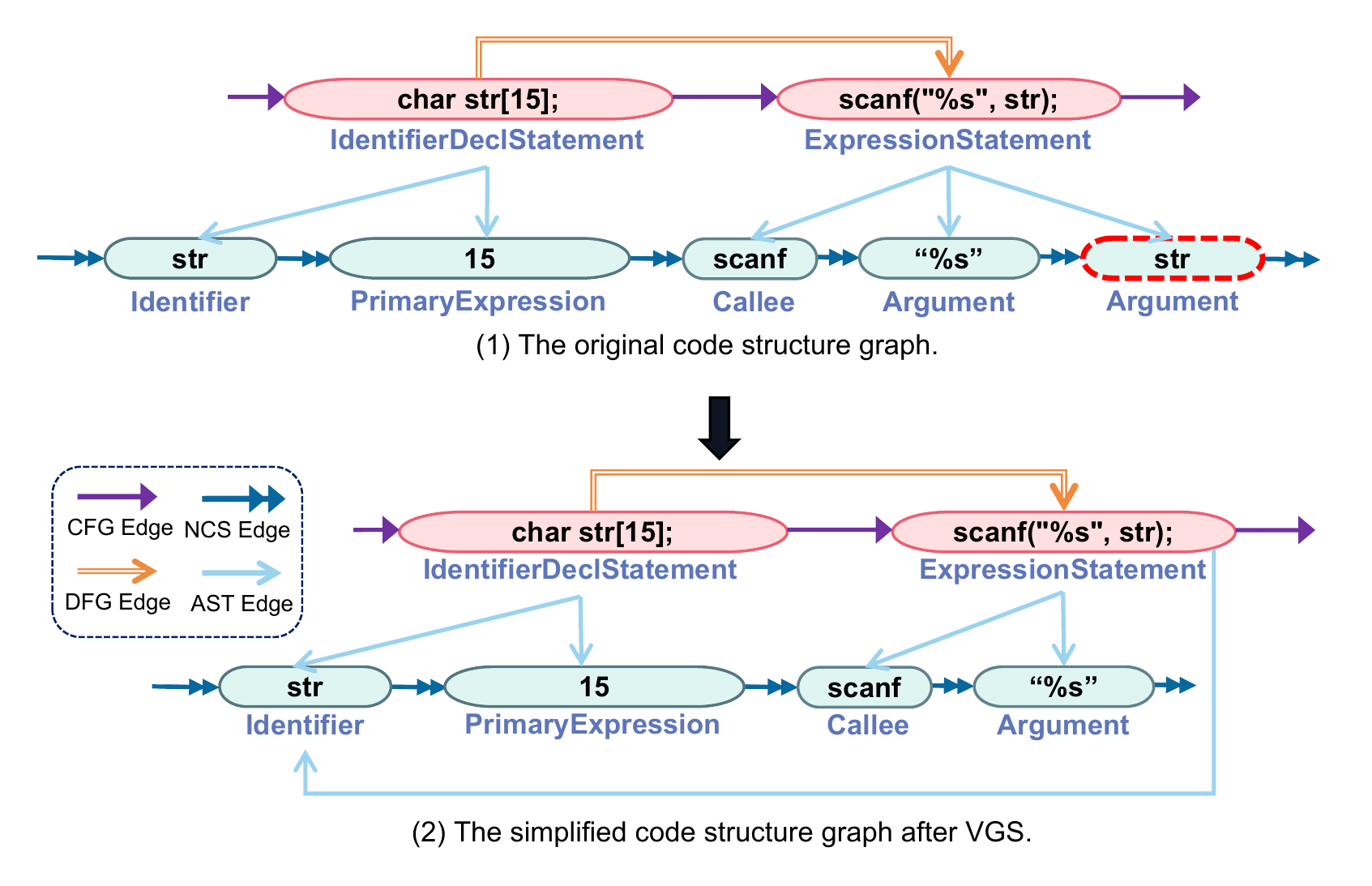}
\caption{An example illustrating variable-based graph simplification. The node with red dashed border in the
original code structure graph (1) is merged to the previous node ``\textit{str}'', producing a VGS-based simplified code structure graph (2).}
\label{cs_2}
\end{figure}

\subsubsection{Variable-based Graph Simplification}
{The variable-based graph simplification (VGS) method aims at merging leaf nodes according to the node variables. Specifically, the VGS method}
merges the {nodes with repeated variables into one node}
in the code structure graph.
It only applies to the leaf nodes of AST and does not change the {hierarchical} parent-child information~\cite{DBLP:conf/aaai/WangL21a}.
{Besides,} since the merged variable node has more than one parent node, it can aggregate information from different statements simultaneously. {This can enhance the node representations and facilitate the global graph representation learning.}
Figure 4 presents an example of variable-based graph simplification. The variable ``\textit{str}'' appears in {the child nodes of} both ``\textit{char str[15]; }'' and ``\textit{scanf(''\%s'',str);}''{, thereby} we merge the two ``\textit{str}'' {leaf} nodes. 


\subsubsection{The Algorithm {of} Graph Simplification}
The {overall graph simplification {(\final)} process is depicted in}
Algorithm \ref{algorithm1}, {in which TGS and VGS correspond to Lines 2-19 and Lines 20-25, respectively.}
The algorithm takes the original code structure graph as the input, and outputs the simplified code structure graph {based on TGS and VGS}. For TGS, the algorithm performs breadth-first traversal on AST. Each time encountering a pair of adjacent nodes $(u,v)$ conforming to the {merging rules in Table~\ref{type}}
(Line 10), the following operations are performed (Lines 11-12): deleting $v$ and the edges connected with $v$, {and then} adding new edges between $u$ and all the child nodes of $v$. For VGS, the algorithm {first} obtains all
groups
{containing the} same variable, and then merges the nodes with same variable
into one node. {The final simplified code structure graph is fed into the subsequent part of \tool.} 


\begin{algorithm}[!tpb]
    \SetAlgoLined
    \footnotesize
    \SetKwInOut{Input}{Input}
    \SetKwInOut{Output}{Output}
    \SetKwFunction{Simplification}{Simplification (GS)}
    \SetKwProg{Fn}{Function}{:}{}
    \Input{Original Code Structure Graph: $Graph_{Original}$}
    \Output{Simplified Code Structure Graph: {$Graph_{\final}$}} 
    \Fn{{\final}}{
    \tcp{{The procedure of TGS}}  
     $Graph_{TGS} \leftarrow Graph_{Original}$ 
     
      \For{each AST \textbf{T} $\in Graph_{TGS}$}{ \tcp{Do breadth-first traversal} 
    $Stack \leftarrow \varnothing$ 
    
    Pushing  $\textbf{T}.root$ into $Stack$ 
    
    \While{$Stack \neq \varnothing$}{
    $u \leftarrow$ Pop $Stack$ 
    
    $Childen \leftarrow$ Get all the child nodes of $u$ 
    
    \For{$v \in Children$}{
    \eIf{{(u.Type, v.Type) conform to rules in Table~\ref{type}}}{
    Delete edge $(u, v)$ and $v$ from $Grpah_{TGS}$; 
    
    Add edges between $u$ and all child nodes of $v$; 
    
    Push all child nodes of $v$ into $Stack$; } { Push $v$ into $Stack$; }}}}
    \tcp{{The procedure of VGS}}
    $Graph_{\final} \leftarrow Graph_{TGS}$ 
    
    $leaf\_nodes \leftarrow$ get all the leaf nodes of AST in $Graph_{\final}$
    
    $same\_variable \leftarrow$ get all the node groups with the same variable in $leaf\_nodes$ 
    
    \For{each $group \in same\_variable$ }{Merging all the nodes in the $group$ into one variable node}
    
    }
    \Return $Graph_{\final}$
\caption{Graph simplification}
\label{algorithm1}
\end{algorithm}
\subsection{Enhanced Graph Representation Learning}
To enhance graph representation learning,
we design two modules: the \textbf{edge-aware graph convolutional network module} for fusing heterogeneous edge information into node representations, and the \textbf{kernel-scaled representation module} which scales up the convolutional kernel size to capture the relations between distant nodes in the graph.

\subsubsection{Edge-Aware Graph Convolutional Network Module} \label{Edge-aware}
The edge-aware graph convolutional network (EA-GCN), as shown in the left of Figure~\ref{architecture} (B), is proposed to exploit different edge types, \eg, AST and CFG, in the directed simplified graph for enhancing node representations. EA-GCN module first computes node vectors by weighting edges of different types separately, and then enhances node representations based on multi-head attention.

We denote the input simplified code structure graph
as $G(\mathcal{V}, \mathcal{E}, \mathcal{R})$. $\mathcal{V}$ is the set of nodes in the graph and each node $v_i \in \mathcal{V}$ is initialized as $h_{i}^{0}\in \mathbb{R}^{d}$ by word2vec \cite{wv}, where $d$ is the vector dimension.
$\mathcal{E}$ represents the set of edges and  $(v_i, \beta, v_j) \in \mathcal{E}$ denotes a labeled edge, where $\beta \in \mathcal{R}$ indicates the edge type and $\mathcal{R}$ is the set of edge types.

During the message passing, we propose to incorporate multiple edge information into node embeddings. Specifically, for node $v_i$ with edge type $\beta$ in the graph, we compute the edge weight $W_{\beta}^{l}$ in the $l$-th layer as:

\begin{equation}
W_{\beta}^{l} = a_{\beta}^{l}V^{l},
\label{rgcn3}
\end{equation} 

\noindent where $a_{\beta}^{l}$ is a learnable weight specific to the edge type $\beta$, and $V^{l} \in R^{d \times d}$ represents a linear transformation. We then employ the edge weight to update the representation of the node $v_i$ based on the following propagation mechanism:

\begin{equation}
{h}_{i}^{l} = \sigma \left ( \sum_{\beta \in \mathcal{E}}\sum_{j\in N_{i}^{\beta}}\frac{1}{c_{i,\beta}} W_{\beta}^{l}{h}_{j}^{l-1} + W_{0}^{l}{h}_{i}^{l-1}\right ), 
\label{rgcn2}
\end{equation}
where ${h}_{i}^{l}$ is the hidden state of node $v_{i}$, $N_{i}^{ \beta }$ denotes the set of indices of neighborhood nodes with the edge type $\beta$, and $c_{i, \beta }$ is the number of neighboring nodes.

To further exploit the heterogeneous edge information in the graph, we propose to adopt multi-head attention mechanism~\cite{DBLP:conf/nips/VaswaniSPUJGKP17}.
The attention mechanism is capable of controlling how much different edge information contributes to the node representation. The attention score $w_{i\rightarrow j}^{k,l}$ for the edge from source node $v_i$ to destination node $v_j$ is computed as below.

\begin{equation}
w_{i\rightarrow j}^{k,l} = softmax_{j}\left ( \frac{{h}_{i}^{k,l}\cdot {h}_{j}^{k,l}}{\sqrt{d_{k}}} \right ),
\label{transformer1}
\end{equation}

\noindent where $k$ denotes the attention head number, and ${h}_{i}^{k,l}$ and ${h}_{j}^{k,l}$ denote the representations of $v_i$ and $v_j$, respectively. The node representation ${h}_{i}^{l}$ is enhanced by aggregating the attention scores of the node edges:

\begin{equation}
{A}_{i}^{l} = Concat_{k=1}^{P}(\sum_{j \in N_{i}}w_{i\rightarrow j}^{k,l}{h}_{j}^{k,l})W_{h}^{l} + h_{i}^{l-1},
\label{transformer2}
\end{equation}

\begin{equation}
h_{i}^{l} = W_{2}^{l} \cdot \sigma \left( W_{1}^{l}{A}_{i}^{l} \right )+{A}_{i}^{l},
\label{transformer4}
\end{equation}

\noindent where $N_{i}$ denotes the set of neighboring nodes of the node $v_i$ and $P$ is the number of attention heads.
The $Concat(\cdot)$ denotes the combination of representations of different heads, and $\sigma(\cdot)$ is the activation function.
$W_{1}^{l}$ and $W_{2}^{l}$ 
are linear transformation layers with bias terms. We finally compute the edge-enhanced node representation matrix of the whole graph as $H_{i}^{l} = \left \{ {h_{k}^l}\right\}_{q=1}  ^{\left|\mathcal{V} \right|}$. $\left|\mathcal{V}\right|$ denotes the number of nodes in the graph.

\subsubsection{Kernel-scaled Representation Module}\label{Kernel-scaled}

The module aims at learning the global information of graphs by explicitly capturing the relations between distant nodes. Specifically, the kernel-scaled representation module involves two convolution kernels with different scales, \ie, one with a large kernel and the other with a small kernel. The large kernel and small kernel are designed to focus on the relations between distant nodes and neighborhood nodes, respectively.

Given the edge-enhanced node representation matrix $H_{i}^{l}$, the module conducts kernel-scaled convolution, in which the large and small kernel sizes are denoted as $N$ and $M$ $(M<N)$, respectively. The convolution with different kernel scales is conducted in parallel, and the output based on the double-branch convolution is defined as:

\begin{equation}
K^{out} = BN \left( H_{i}^{l}  \ast W^{L}, \mu^{L} \right) + BN(H_{i}^{l}  \ast W^{S}, \mu^{S}),
\label{conv_eq}
\end{equation}

\noindent where $\ast$ is the convolution operator. $W^L \in R^{C_{out} \times C_{in} \times N}$ and $W^S \in R^{C_{out} \times C_{in} \times M}$ denote the large convolution kernel and small convolution kernel, respectively, where $C_{in}$ and $C_{out}$ are the input and output channels of the convolution. The results of the two branches are summed after passing through a batch normalization (BN) layer \cite{ioffe2015batch}. $\mu^{L}$ is the parameter of the BN layer following the large convolution kernel and $\mu^{S}$ is for the other branch.
Finally, we use two fully-connected (FC) layers and a softmax function to perform binary classification (vulnerable or non-vulnerable).

\section{Experimental Setup}
\label{sec:experimentalsetup}

\subsection{Research Questions}

In order to evaluate \tool, we answer the following research questions:

\begin{enumerate}[label=\bfseries RQ\arabic*:,leftmargin=.5in]
    \item How effective is \tool in vulnerability detection?
    \item How does graph simplification contribute to the performance of \tool?

    \item How does enhanced graph representation learning contribute to the performance of \tool?

    \item What is the {influence of hyper-parameters on the performance of \tool?}
\end{enumerate}




\subsection{Datasets}
To investigate the effectiveness of \tool, following \ivd \cite{IVDETECT}, we adopt three vulnerability datasets in our study, including FFMPeg+Qemu \cite{devign}, Reveal \cite{reveal}, and Fan \et. \cite{fan}. {The statistics of the experimental datasets are illustrated in Table~\ref{dataset}.}
The FFMPeg+Qemu dataset provided by Devign~\cite{devign} is manually-labeled and comes from two open-source C projects. It contains about 10k vulnerable entries and 12k non-vulnerable entries. 
The Reveal dataset~\cite{reveal} is collected from two open-source projects: Linux Debian Kernel and Chromium. It consists of about 2k vulnerable entries and 20k non-vulnerable entries. 
Fan \et.~\cite{fan} collected from over 300 open-source C/C++ GitHub projects, which cover 91 different vulnerability types from 2002 to 2019 in the Common Vulnerabilities and Exposures (CVE) database. The dataset consists of approximately 10k vulnerable entries and 177k non-vulnerable entries. 
\begin{table}[t]
\centering
\setlength{\tabcolsep}{1.1mm}
\renewcommand{\arraystretch}{1.1}

\caption{Statistics of the datasets.}
\begin{tabular}{c|r|r|r|r}
\toprule
Dataset & \multicolumn{1}{c|}{Samples} &  \multicolumn{1}{c|}{Vul}  & \multicolumn{1}{c|}{Non-vul} & \multicolumn{1}{c}{Vul Ratio(\%)} \\

\midrule
FFMPeg+Qemu~\cite{devign}  & 22,361   & 10,067 & 12,294   & 45.02  
\\
Reveal~\cite{reveal}  & 18,169   & 1,664  & 16,505   & 9.16          \\
Fan \et.~\cite{fan}     & 179,299  & 10,547 & 168,752  & 5.88        \\
\bottomrule

\end{tabular}
\label{dataset}
\end{table}

\subsection{Baseline Methods}

In our evaluation, we compare \tool with three state-of-the-art graph-based methods~\cite{devign, reveal, IVDETECT} and three token-based methods~\cite{vuldeepecker, russell, sysevr}.

\textbf{(1) Devign} \cite{devign}: Devign builds a joint graph containing AST, CFG, DFG and NCS, and uses GGNN for vulnerability detection.

\textbf{(2) Reveal} \cite{reveal}: Reveal divides code vulnerability detection into two phases: feature extraction and training. It leverages GGNN to extract features. 

\textbf{(3) \ivd} \cite{IVDETECT}: \ivd constructs PDG and utilizes GCN to learn the graph representation for vulnerability detection. 

\textbf{(4) VulDeePecker} \cite{vuldeepecker}: VulDeePecker embeds source code and the data/control dependencies in a program slice through word2vec. It then feeds the embedding into a bidirectional LSTM-based neural network with an attention mechanism for vulnerability detection.

\textbf{(5) Russell \et.} \cite{russell}: Russell \et. label the source code and embed it into the corresponding matrix. They adopt convolutional neural networks, integrated learning, and random forest classifier for code vulnerability detection.

\textbf{(6) SySeVR} \cite{sysevr}: SySeVR uses code statements, program dependencies, and program slicing as features, and utilizes a bidirectional recurrent neural network to detect vulnerable code snippets.

\subsection{Implementation Details}

To ensure the fairness of the experiments, we use the same data splitting for all approaches. We randomly partition the dataset into disjoint training, validation, and test sets with the ratio of 8:1:1. {We reproduce all the baseline models except Devign based on the publicly released source code and {use the same} hyperparameter settings} {as} described in their original paper. {Since Devign itself does not provide the source code, we reproduce the baseline based on
the implementation provided by Reveal~\cite{reveal}. } 

{Following the work~\cite{reveal}, we create the code structure graph based on the results parsed by Joern~\cite{yamaguchi2014modeling}. The preprocessing steps used to initialize the embedding vectors of each node are as follows: (1) we first tokenize the code in this node as a token sequence, with punctuations kept (2) we initialize the embedding of each token through word2vec~\cite{wv}, without removal of any tokens (3) we compute the average of the token embeddings as the initial embedding of this node. }
In the embedding layer, the {dimension of} the initial node representation $d_{k}$ is set as
100. We adopt RAdam \cite{RAdam} optimizer with a learning rate at 0.0001. The batch size is set to 64. The large kernel $M$ and small kernel $N$ are set to 11 and 3 respectively. The number of EA-GCN layers is 2 and the number of the attention head $P$ is 10.
We train our model for a maximum of 100 epochs on a server with NVIDIA GeForce RTX 3090 with 20-epoch patience for early stopping.

\subsection{Performance Metrics}

We use the following four widely-used performance metrics in our evaluation:

\textbf{Precision:} $Precision = \frac{TP}{TP+FP}$. Precision is the proportion of relevant instances among those retrieved. $TP$ the number of true positives and $FP$ the number of false positives.

\textbf{Recall:} $Recall = \frac{TP}{TP+FN}$. Recall is the proportion of relevant instances retrieved. $TP$ the number of true positives and $FN$ is the number of false negatives.

\textbf{F1 score:} $F1\ score = 2 \times \frac{Precision\times  Recall}{Precision + Recall}$. F1 score is the geometric mean of precision and recall and indicates balance between them.

\textbf{Accuracy:} $Accuracy = \frac{TP+TN}{TP+TN+FN+FP}$. Accuracy is the proportion of correctly classified instances to all instances.
$TN$ is the number of true negatives and $TP+TN+FN+FP$ represents the number of all samples.

\label{sec:evaluation}

\section{Results}
\label{sec:results}

\subsection{RQ1. {Effectiveness of \tool}}

\begin{table*}[h]
\centering

\setlength{\tabcolsep}{1.2mm}
\renewcommand{\arraystretch}{1.1}

\caption{
Comparison results between \tool and the baselines on the three datasets. ``-'' means that the baseline does not apply to the dataset in this scenario. 
The best result for each metric is highlighted in bold. 
The cells in grey represent the performance of the top-3 best methods in each metric, with darker colors representing better performance.} 
\resizebox{.97\textwidth}{!}{
\begin{tabular}{l|cccc|cccc|cccc}
\toprule
\diagbox{Metrics(\%) }{Dataset} & \multicolumn{4}{c|}{FFMPeg+Qemu \cite{devign}}        & \multicolumn{4}{c|}{Reveal \cite{reveal}}             & \multicolumn{4}{c}{Fan \et. \cite{fan}}                \\
\midrule
Baseline                         & Accuracy & Precision & Recall & F1 score     & Accuracy & Precision & Recall & F1 score    & Accuracy & Precision & Recall & F1 score    \\
\midrule
VulDeePecker                    & 49.61   & 46.05    & 32.55 & 38.14 & 76.37   & 21.13    & 13.10 & 16.17 & 81.19   & \cellcolor{gray!70}\textbf{38.44}    & 12.75 & 19.15 \\

Russell \et.                  & \cellcolor{gray!20}57.60   & \cellcolor{gray!20}54.76    & 40.72 & 46.71 & 68.51   & 16.21    &\cellcolor{gray!20} 52.68 & 24.79 & 86.85   & 14.86    & \cellcolor{gray!20}{26.97} & 19.17 \\

SySeVR                          & 47.85   & 46.06    & 58.81 & 51.66 & 74.33   & \cellcolor{gray!40}40.07  & 24.94 & 30.74   & \cellcolor{gray!20}90.10   & \cellcolor{gray!40}30.91   & 14.08 & 19.34 \\

Devign                          & 56.89   & 52.50    & \cellcolor{gray!20} 64.67 & \cellcolor{gray!20} 57.95 &\cellcolor{gray!40} 87.49   & \cellcolor{gray!20}31.55    & 36.65 &\cellcolor{gray!20} 33.91 & \cellcolor{gray!40}92.78   &\cellcolor{gray!20} 30.61   & 15.96   & \cellcolor{gray!20}20.98 \\

Reveal                          & \cellcolor{gray!40}61.07   & \cellcolor{gray!40}55.50    &\cellcolor{gray!40} 70.70 &\cellcolor{gray!40} 62.19 &\cellcolor{gray!20}  81.77   &\cellcolor{gray!20}  31.55    & \cellcolor{gray!70}\textbf{61.14}&\cellcolor{gray!40} 41.62   & 87.14   & 17.22   &\cellcolor{gray!40} 34.04 & \cellcolor{gray!40}22.87 \\

IVDetect                        & 57.26   & 52.37    & 57.55 & 54.84 & -   & -         & -      & -      & -   & -         & -      & -      \\
\midrule
\tool
& \textbf{\cellcolor{gray!70}62.16}   & \cellcolor{gray!70}\textbf{55.64}    & \cellcolor{gray!70}\textbf{83.99} &\cellcolor{gray!70} \textbf{66.94} &\cellcolor{gray!70}\textbf{92.71}   & \cellcolor{gray!70} \textbf{51.06}    & \cellcolor{gray!40} 46.15 & \cellcolor{gray!70}\textbf{48.48} & \cellcolor{gray!70}\textbf{93.14}   & 29.98    & \cellcolor{gray!70}\textbf{34.58} & \cellcolor{gray!70}\textbf{32.11}\\
\bottomrule
\end{tabular}}
\label{table_rq1}
\end{table*}

To answer the first question, we compare \tool{} with the six baseline methods
on the three datasets. The results are illustrated in Table \ref{table_rq1}. Based on the results, we achieve the following observations.

\textbf{Graph-based methods perform better than token-based methods: }We observe that the three graph-based methods Devign, Reveal and IVDetect
{show superior performance than the}
three token-based methods in terms of accuracy and F1 score. {For example, the cells with grey colors which indicate the top-3 best methods generally appear in the bottom rows of the table.} {The results demonstrate that}
graph-based methods can better capture the code structural information,
which {is beneficial for} 
improving the performance of code vulnerability detection.  

\textbf{The proposed methods consistently outperform all the baseline methods:}
We observe that \tool outperforms all the baseline methods on the three datasets in terms of accuracy and F1 score. 
Specifically, \tool{} achieves 4.75\%, 6.86\%, and 9.24\% absolute improvement in F1 score over the best baseline method on the three datasets, respectively. 
The corresponding relative improvements are 7.63\%, 16.48\%, and 40.40\%. When considering all the four performance metrics regarding the three datasets (12 combination cases altogether), \tool{} has the best performance in 10 out of the 12 cases, and top-3 performance in 11 out of the 12 cases. {We also make an additional comparison with LineVul~\cite{fu2022linevul} on the FFMPeg+Qemu and Reveal datasets. On these two datasets, LineVul achieves the F1-score of 56.54\% and 44.79\%, respectively; while our proposed model achieves 66.94\% and 48.48\%, respectively.}




 \begin{tcolorbox}
 \textbf{Answer to RQ1:} \tool outperforms all the baseline methods in terms of accuracy and F1 score. 
 In particular, \tool{} achieves 7.63\%, 16.48\%, and 40.40\% improvements in F1 score over the best baseline method on the three datasets, respectively. 
 \end{tcolorbox}

\subsection{RQ2. Effectiveness of Graph Simplification}
To answer the research question, we first 
explore the contribution of the graph simplification to the performance of \tool{} and the effectiveness of different simplification methods. We then {analyze} the graph simplification rate (\ie, the ratio of reduced nodes and edges in code structure graphs) and the node distance (\ie, average node distance and maximum node distance) in the simplified code structure graph. 

\subsubsection{Impact on Model Performance}
\label{sec:graphsimplicationcontribution}
To evaluate the contribution of the graph simplification and the effectiveness of our proposed simplification methods, we perform an ablation study on the three datasets. 
In particular, we compare the performance of four versions of \tool{}: without graph simplification (denoted as $Graph_{Original}$), with only TGS 
(denoted as $Graph_{\namei}$), with only VGS 
($Graph_{\nameii}$), and with both graph simplification methods ($Graph_{\final}$, the default \tool{}){, with results shown in Table~\ref{table_rq2}. The best result for each metric of each dataset is highlighted in bold.}

\begin{table*}[h!]
\centering

\setlength{\tabcolsep}{1.1mm}
\renewcommand{\arraystretch}{1.1}
\caption{Effectiveness of graph simplification. 
``$Graph_{Original}$'' represents the performance of \tool on the original code structure graphs.
``$Graph_{\namei}$'' and ``$Graph_{\nameii}$'' represent \tool{}'s performance on the type-based simplification strategy (\namei) and variable-based simplification strategy (\nameii), respectively. ``$Graph_{\final}$'' denotes the performance of \tool with both \namei and \nameii.}
\resizebox{.97\textwidth}{!}{
\begin{tabular}{l|cccc|cccc|cccc}
\toprule
\diagbox{Metrics(\%) }{Dataset} & \multicolumn{4}{c|}{FFMPeg+Qemu \cite{devign}}        & \multicolumn{4}{c|}{Reveal \cite{reveal}}             & \multicolumn{4}{c}{Fan \et. \cite{fan}}                    \\
\midrule
Baseline                       & Accuracy & Precision & Recall & F1 score    & Accuracy & Precision & Recall & F1 score    & Accuracy & Precision & Recall & F1 score     \\
\midrule
$Graph_{Original}$                & 58.32   & 52.91    & 67.87 & 59.47 & 87.99   & 31.22    & \textbf{66.29} & 42.45 & 87.87   & 16.48    & \textbf{42.88} & 23.81 \\

$Graph_{\namei}$               & 60.82   & 54.85    & 75.79 & 63.64 & 89.08   & 34.55    & 61.96 & 44.36 & 89.33  & 19.48    & 42.07 & 26.63 \\

$Graph_{\nameii}$                & 59.66   & 53.59    & 81.82 & 64.76 & 89.68   & 37.33    & 57.14 & 45.16 & 91.69  & 23.01    & 33.87 & 27.40\\

\midrule
$Graph_{\final}$
         & \textbf{62.16}   & \textbf{55.64}    & \textbf{83.99} & \textbf{66.94} 
 &\textbf{92.71}   & \textbf{51.06}    &46.15 & \textbf{48.48} & \textbf{93.14}   & \textbf{29.98 }   & 34.58 & \textbf{32.11}\\
\bottomrule

\end{tabular}}

\label{table_rq2}
\end{table*}
\begin{figure}
    \centering
    \subfloat[Average node distance]{
    \begin{minipage}[t]{0.46\linewidth}
    \centering
    \includegraphics[width=1.0\textwidth]{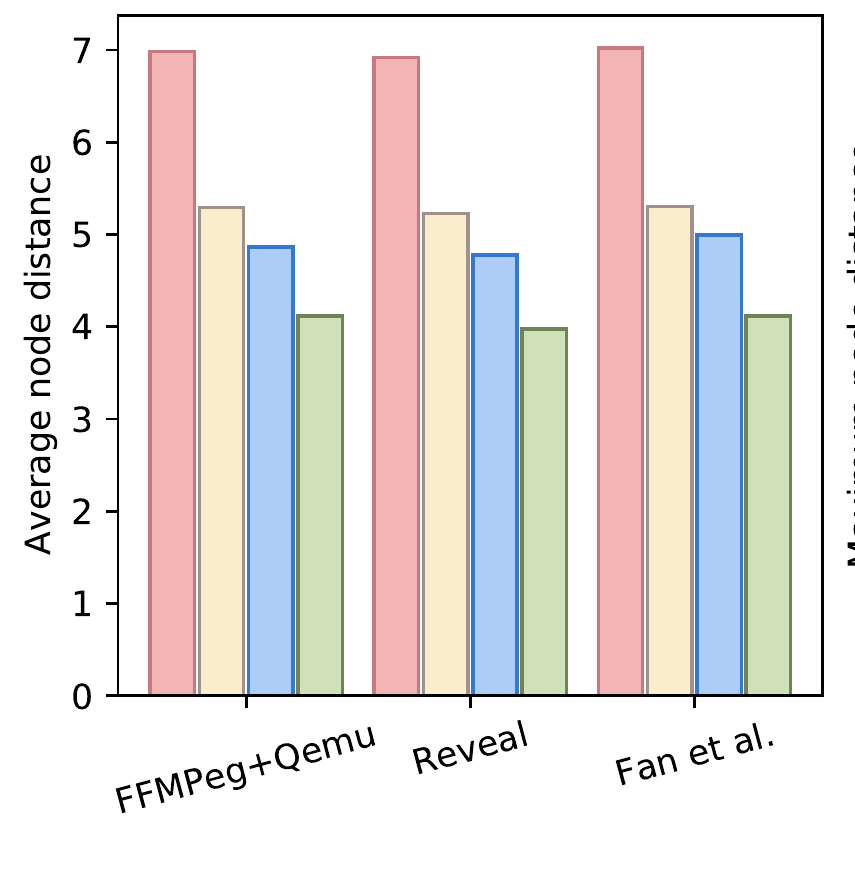}
    \label{Avg}
    \end{minipage}%
    }
    \hspace{1mm}
    \vspace{-1mm}
    \subfloat[Maximum node distance]{
    \begin{minipage}[t]{0.46\linewidth}
    \centering
    \includegraphics[width=1.025\textwidth]{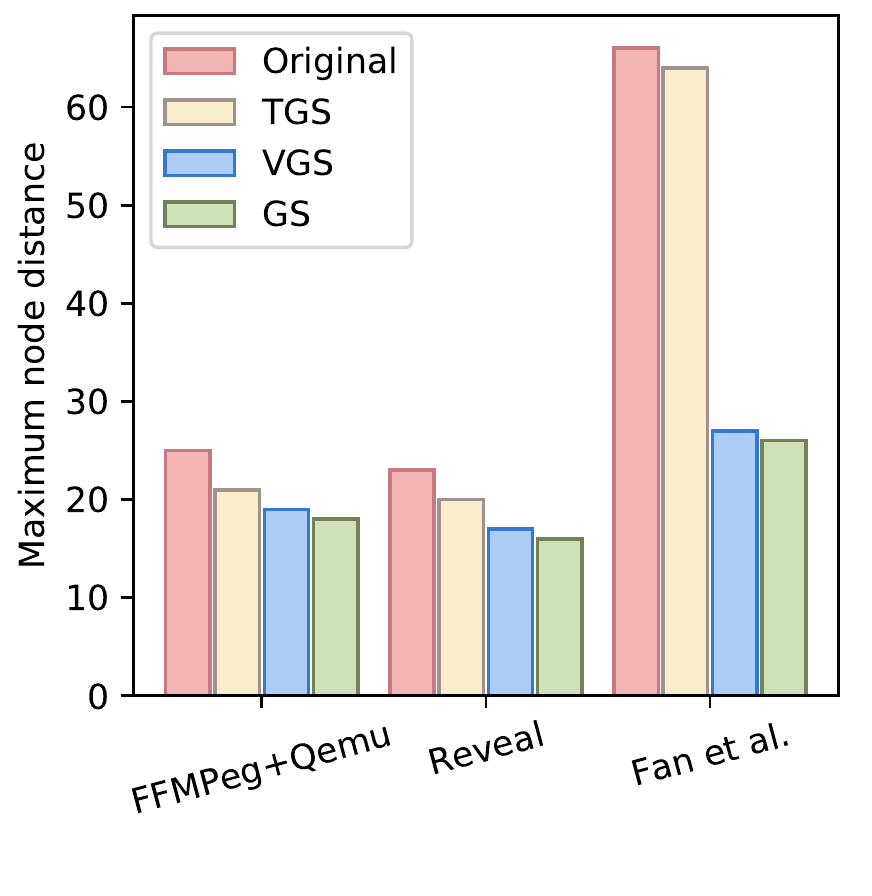}
    
    \label{Max}
    \end{minipage}
    }
\centering
\caption{Average node distance (a) and maximum node distance (b) on the test set
 of FFMPeg+Qemu, Reveal, and Fan \et. 
The node distance is measured by the length of the shortest path between two nodes in a  graph.
}
\label{distance}
\end{figure}

Compared with $Graph_{Original}$, $Graph_{\final}$ improves the accuracy by 6.58\%, 5.36\%, and 6.00\% on three datasets, and improves the F1 score by 12.56\%, 14.20\%, and 34.86\% respectively. Both $Graph_{\namei}$ and $Graph_{\nameii}$ outperform $Graph_{Original}$ in terms of accuracy and F1 score on three datasets, which demonstrates our proposed graph simplification methods can {contribute to}
the performance of \tool{}. {In addition, $Graph_{\nameii}$ slightly outperforms $Graph_{\namei}$}. {Compared with $Graph_{\namei}$, $Graph_{\nameii}$ improves the accuracy by 0.75\% and F1 score by 2.00\% on three datasets on average. 
The results indicate}
that VGS contributes more to the
graph representation learning than TGS. 

\subsubsection{Analysis of Code Structure Graph Size} 
\label{sec:simplicationrate}
{We analyze the size of the code structure graph from two perspectives, including node distance (\ie the shortest length between two reachable nodes), node and edge simplification rate (\ie the ratio of reduced nodes and edges).}
    

\textbf{Node Distance:} As shown in Figure~\ref{distance}, we count the length of the shortest path between any two reachable nodes as node distance {during calculating}
the average node distance and maximum node distance. Due to the large computational complexity and limited resource, we only calculate the distances on the test sets for the three datasets. {Compared with the original code structure graph, the average node distance and maximum node distance on three datasets drop by 41.65\% and 39.68\%, respectively, which indicates that our proposed simplification methods are helpful to reduce the node distance.} 
We also notice that \nameii can reduce average node distance and maximum node distance more than \namei, { 30.05\% v.s. 24.38\ in terms of the average node distance. The results can explain the relatively better performance of $Graph_{\nameii}$ than $Graph_{\namei}$ in vulnerability detection (as illustrated in Table~\ref{table_rq2}).} 

\begin{figure}[t]
   
	\centering

	\includegraphics[width=0.5\textwidth]{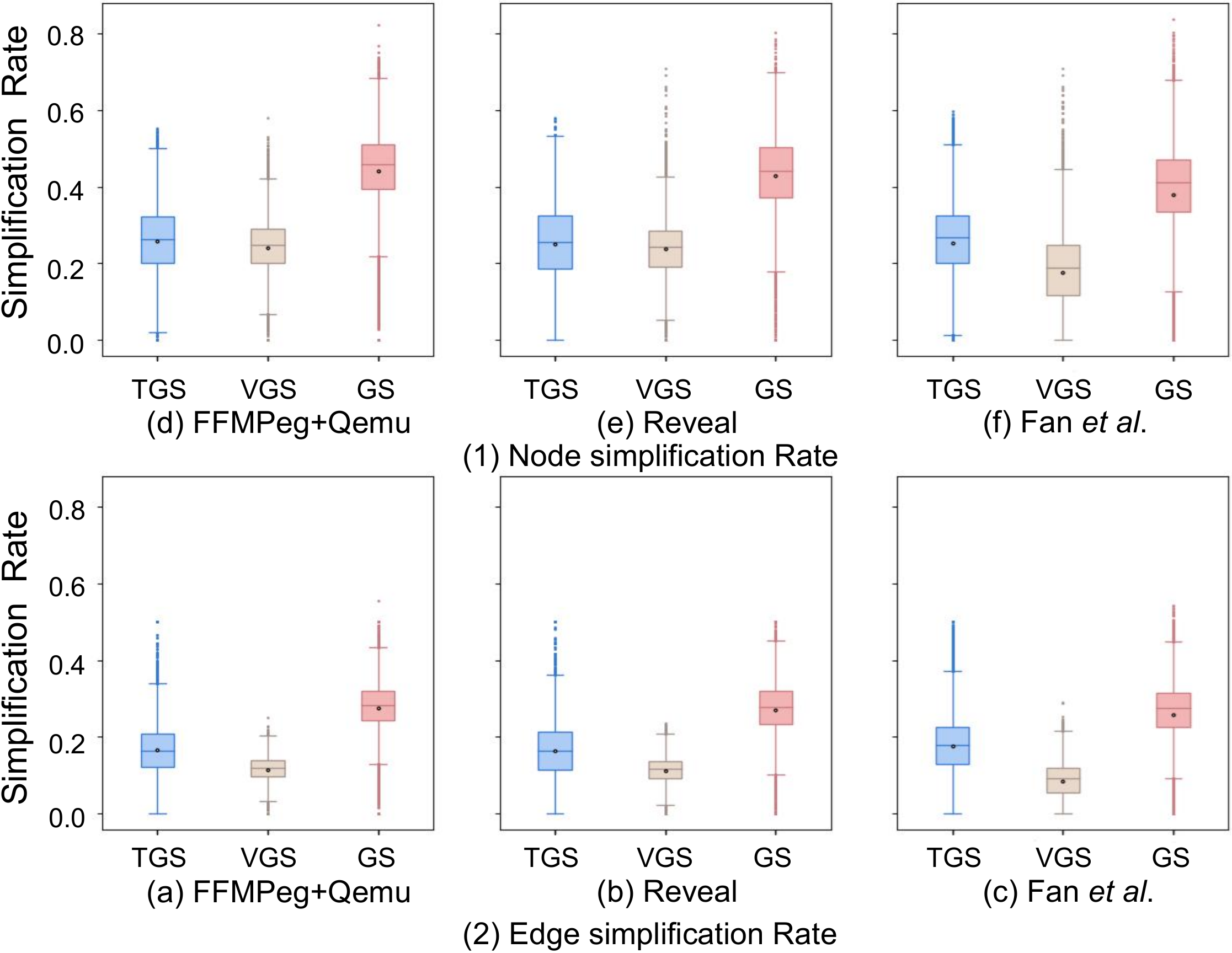}
	\caption{Edge simplification rate and node simplification rate on three datasets. 
	}
    \label{cs_result}
    \vspace{-6pt}
\end{figure}

\textbf{Node and Edge Simplification Rate:}
Figure~\ref{cs_result} shows the box-whisker plot of node and edge simplification rate on each dataset. The average simplification rate is {computed by taking} the average of the simplification rates {for all the graphs.}
The overall distribution of simplification rates shows that both \namei and \nameii can obviously reduce the node size and edge size of the code structure graph. With the same simplification method, the node simplification rate is higher than the edge simplification rate. Specifically, the average node simplification rate and edge simplification rate of the final simplified code structure graph on three datasets are 41.64\% and 16.79\%, respectively. Comparing the simplification rates of \namei and \nameii, we find that \namei can reduce the code structure graph size to a greater extent, which may be attributed to that \nameii only applies to the leaf nodes of AST.

\begin{tcolorbox}
\textbf{Answer to RQ2:} Graph simplification contributes significantly to the performance of \tool{}, with an F1 score improvement of 
12.56\%, 14.20\%, and 34.86\% on the three datasets, respectively.
The average node, edge, and distance simplification rate is 
41.64\%, 16.79\%, and 41.65\%, respectively. 
\end{tcolorbox}

\subsection{RQ3. Effectiveness of Enhanced Graph Representation Learning}

To answer this research question, we explore the impact of the enhanced graph representation learning on \tool. {Specifically, we study the two involved modules including the edge-aware graph convolutional network module (EA-GCN) and the \ksr(\ksrab).}

\subsubsection{Edge-Aware Graph Convolutional Network Module}
To explore the contribution of EA-GCN module, we create a variant of \tool without EA-GCN module that directly feeds the simplified code structure graph into the \ksrab module. The other settings of this variant are consistent with \tool.
As shown in Table~\ref{table_rq3}, EA-GCN module improves the {the performance of \tool}
on all datasets, \eg, 4.46\% in FFMPeg+Qemu, 30.08\% in Reveal, and 48.66\% in Fan \et, {in terms of the F1 score.} 
{The results} indicate that the EA-GCN
module {focusing on the heterogeneous edge information in the graph can promote the graph representation learning and benefit the performance of vulnerability detection.}


\begin{table}[t]
\centering

\setlength{\tabcolsep}{1.2mm}
\renewcommand{\arraystretch}{1.1}
\caption{Impact of the EA-GCN module and the \ksrab module on the performance of \tool. 
}
\resizebox{.47\textwidth}{!}{
\begin{tabular}{c|c|cccc}
\toprule
\multicolumn{1}{c|}{Dataset} & Module      & Accuracy & Precision & Recall & F1 score \\\midrule
\multirow{3}{*}{FFMPeg+Qemu}     

                            & w/o EA-GCN & 57.13   & 51.60    & \textbf{84.53} & 64.08   \\
                            
                            & w/o \ksrab   & 55.89   & 50.90    & 70.65 & 59.17   \\
                            & \tool     & \textbf{62.16}   & \textbf{55.64}    & 83.99 & \textbf{66.94}   \\
                            \midrule
\multirow{3}{*}{Reveal}     
                            & w/o EA-GCN & 85.57   & 27.52    & 57.69 & 37.27   \\
                            & w/o \ksrab  & 79.14   & 20.06    & \textbf{60.58} & 30.14   \\
                            & \tool    & \textbf{92.71}   & \textbf{51.06}    & 46.15 & \textbf{48.48}   \\
                            \midrule
\multirow{3}{*}{Fan \et.}  
                            & w/o EA-GCN & 89.79   & 16.73    & 30.48 & 21.60   \\
                            & w/o \ksrab   & 88.97   & 18.37    & \textbf{39.62} & 25.10  \\
                            & \tool     & \textbf{93.14}   & \textbf{29.98}    & 34.58 & \textbf{32.11}   \\
\bottomrule
\end{tabular}}

\label{table_rq3}
\end{table}
\begin{table*}[h]
\centering

\setlength{\tabcolsep}{1.2mm}
\renewcommand{\arraystretch}{1.1}
\caption{Comparison between the EA-GCN module with other GNN models. 
}
\resizebox{.97\textwidth}{!}{
\begin{tabular}{l|cccc|cccc|cccc}
\toprule
\diagbox{Metrics(\%) }{Dataset} & \multicolumn{4}{c|}{FFMPeg+Qemu \cite{devign}}        & \multicolumn{4}{c|}{Reveal \cite{reveal}}             & \multicolumn{4}{c}{Fan \et. \cite{fan}}                 \\
\midrule
Baseline                        & Accuracy & Precision & Recall & F1 score    & Accuracy & Precision & Recall & F1 score     & Accuracy & Precision & Recall & F1 score \\ 
\midrule
GGNN     & 55.69 & 50.66 & 78.27 & 61.51 & 89.07 & 34.59 & 52.88 & 41.83 & 90.65 & 18.69 & 30.16 & 23.08 \\
GCN          & 60.32 & 54.90  & 68.83 & 61.08 & 88.29 & 32.56 & \textbf{53.85} & 40.58 & 88.31 & 14.84 & 31.79 & 20.23 \\
R-GCN        & 61.40  & \textbf{55.70}  & 71.67 & 62.69 & 89.14 & 34.81 & 52.88 & 41.98 & 90.29 & 16.70  & 27.14 & 20.67 \\\hline
EA-GCN 
& \textbf{62.16}   & \textbf{55.64}    & \textbf{83.99} & \textbf{66.94} 
 &\textbf{92.71}   & \textbf{51.06}    &46.15 & \textbf{48.48} & \textbf{93.14}   & \textbf{29.98 }   & \textbf{34.58} & \textbf{32.11}\\
\bottomrule
\end{tabular}}

\label{table_rq5}
\end{table*}

To further explore the effectiveness of the EA-GCN module
, we replace the module with other existing GNN models, including GCN \cite{kipf2016gcn}, GGNN \cite{Method3}, R-GCN \cite{Method5} and compare them with 
\tool. As shown in Table \ref{table_rq5}, the EA-GCN module 
obtains significant improvements on all databases. In particular, compared to the other best performing GNN methods, the F1 score metric is improved by 6.78\%, 15.48\% and 39.12\% on the three datasets, respectively. {The EA-GCN module learns
different weights for multiple types of edges,
which can benefit
the performance of vulnerability detection.}



\subsubsection{Kernel-Scaled Representation Module}\label{kscr}

To understand how the \ksr works, we also deploy a variant of \tool without this module. The variant directly replaces the KSR module with a fully-connected layer to classify the edge-enhanced node representation matrix learned by EA-GCN.
Table~\ref{table_rq3} shows the performance of the variants.
On all the datasets, the accuracy, F1 score and precision {consistently}
achieve higher value with the addition of KSR module. Specially, compared with the variant without KSR module, the F1 score of \tool is improved by 13.13\%, 57.53\%, and 27.93\% on FFMPeg+Qemu, Reveal, and Fan \et. datasets, respectively. \tool also increases the accuracy by 11.22\%, 17.15\%, and 4.69\%, respectively. The results demonstrate that our proposed KSR module brings a performance improvement in vulnerability detection.

 \begin{tcolorbox}
 \textbf{Answer to RQ3:} 
 The enhanced graph representation learning effectively improves the \tool{} performance.
 The EA-GCN module improves the F1 score of 4.46\%, 30.08\%, and 48.66\% on the FFMPeg+Qemu, Reveal and Fan \et., and the KSR module contributes 13.13\%, 57.53\%, and 27.93\%, respectively.
 \end{tcolorbox}

\subsection{RQ4: Influences of Hyper-parameters on \tool{}}

To answer the research question, we explore the impact of different hyper-parameters, including the sizes of large and small convolution kernel, the layer number of EA-GCN module, and the number of the attention heads.

\subsubsection{{Sizes of Large Kernel and Small Kernel}}
\begin{figure}[t]
	\centering

	\includegraphics[width=0.35\textwidth]{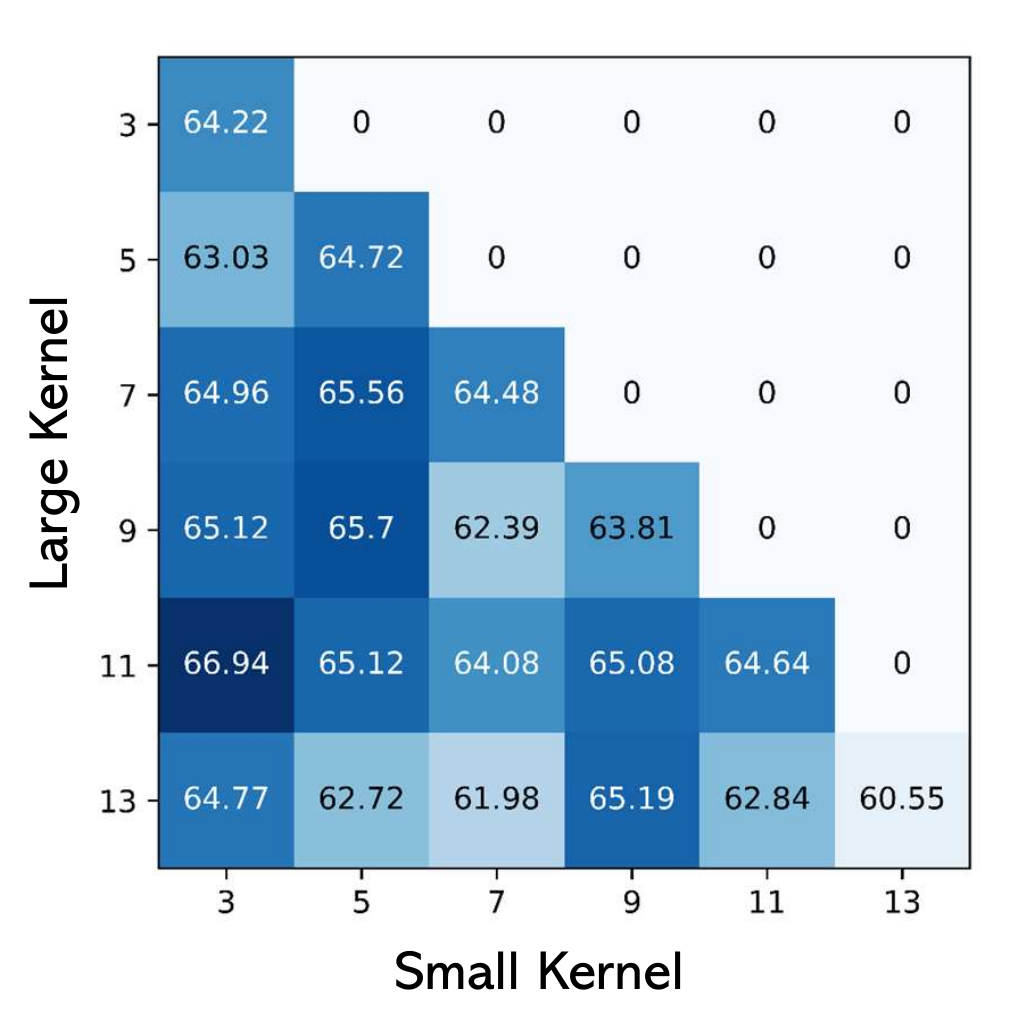}
	\caption{Impact of
	kernel sizes on the F1 score of \tool. {The x-axis and y-axis indicate the small kernel size and large kernel size, respectively.} 
	The darker the color, the better the performance.}
	\label{fi_conv}

\end{figure}

Figure~\ref{fi_conv} shows the F1 score of \tool with different {sizes of} large kernel and small kernel.
\tool achieves the {best performance}
when the {sizes of the} large kernel  and small kernel are set as 11 and 3, respectively.
Two kernels of the same size bring a relatively low F1 value, which indicates that using one type of convolutional kernel does not {well learn the graph representations}.
Besides, due to the limited number of nodes in the simplified code structure graph, the size of the large kernel cannot be increased infinitely and the F1 score starts to decrease {when the kernel size equals to 13.}
In summary, the model benefits from a large kernel to capture the relations between distant nodes, and meanwhile retains the capability of capturing local information in the graph structure from a small kernel. 

\subsubsection{Numbers of EA-GCN layers}
We also explore the effect of the number of EA-GCN layers on the {performance of} \tool.
Table~\ref{hyper} shows the results of EA-GCN with different numbers of layers on FFMPeg+Qemu dataset. 
{As can be seen,} EA-GCN with 2 layers obtains the {highest} accuracy of 62.16\% and F1 score of 66.94\%.  
As the number of layers continues to increase, the accuracy and F1 score decrease. We {suppose}
that the GCN encounters an over-smoothing issue~\cite{bottleneck} {with more layers} and {tends to learn similar embeddings for different nodes},
which leads to {performance degradation.}
\tool exhibits similar trends {on the other datasets}.

\subsubsection{{Numbers of Attention Heads}} Table~\ref{hyper} shows the performance of \tool with different numbers of attention heads. {Generally,} more heads
bring a noticeable improvement in accuracy and F1 score {compared with fewer heads}, which shows that {more heads are helpful for capturing the different edge information}
in the code structure graph.
However, employing more than 10 heads does not further improve performance.

 \begin{tcolorbox}
 \textbf{Answer to RQ4:}
Different settings of hyper-parameters can influence the performance of \tool in vulnerability detection. Our default hyper-parameter settings achieve optimal results.
 \end{tcolorbox}

\begin{table}[t]
\centering
\setlength{\tabcolsep}{1.1mm}
\renewcommand{\arraystretch}{1.1}

\caption{The impact of the number of EA-GCN layers and attention heads on the performance of \tool. 
}
\begin{tabular}{c|cc|c|cc}

\toprule

\multicolumn{1}{c|}{layer number} & Accuracy             & F1 score                  & \multicolumn{1}{|c|}{head number} & Accuracy             & F1 score                  \\
\midrule
1                      & 60.78                & 64.56                & 2                      & 58.92                & 62.43                \\
\textbf{2}             & \textbf{62.16}       & \textbf{66.94}       & 4                      & 61.45                & 63.16                \\
3                      & 61.49                & 61.49                & 5                      & 60.10                & 65.63                \\
4                      & 60.57                & 61.82                & \textbf{10}            & \textbf{62.16}       & \textbf{66.94}       \\
5                      & 55.91                & 59.99                & 20                     & 60.71                & 64.40  \\ 
\bottomrule
\end{tabular}
\label{hyper}
\end{table}

\section{Discussion}
\label{sec:discussion}
\begin{figure}[t]
\centering
\includegraphics[width=0.49\textwidth]{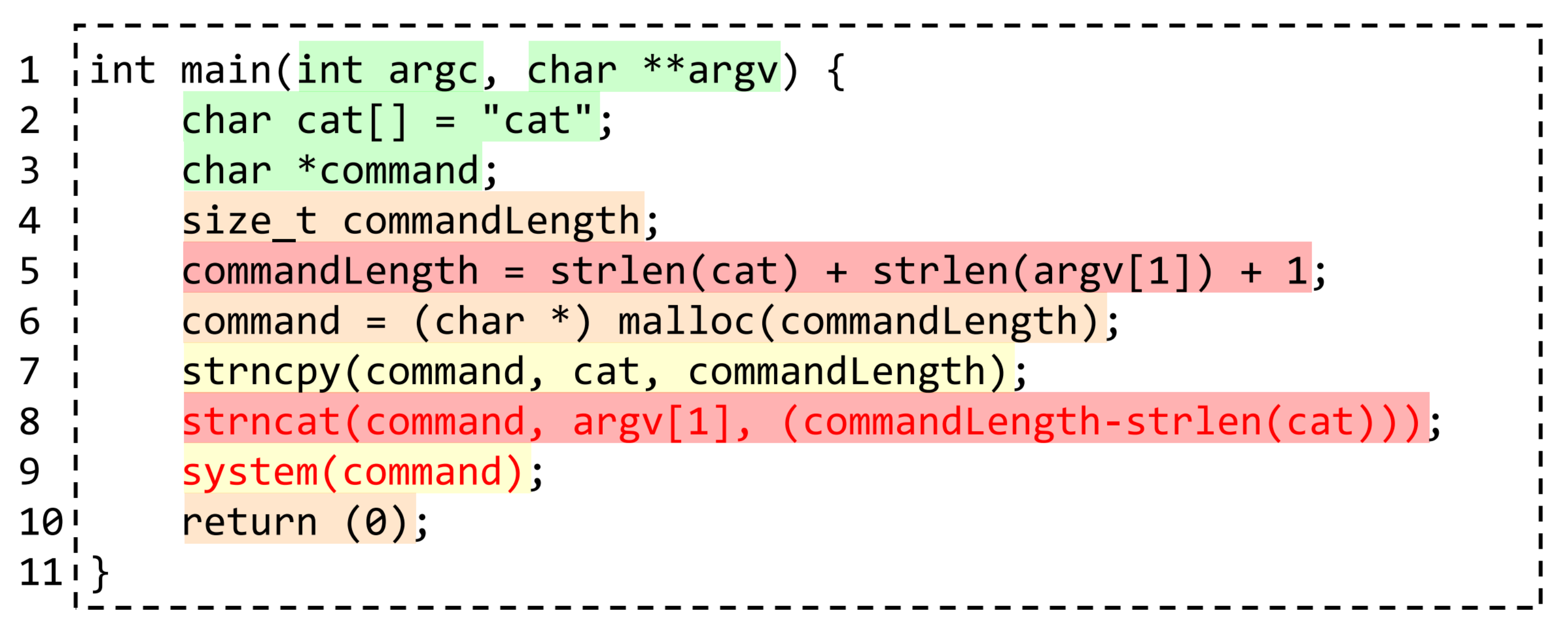}
\vspace{-2ex}
\caption{A heatmap of the CWE-77~\cite{CWE-77} example correctly predicted by \tool. The color-shaded of a statement is drawn according to the `attention' weight. Green-shaded indicates the minimum attention weight. The redder the color, the greater the `attention' weight. Red colored code are vulnerable.}

\label{vul_example}
\end{figure}

\subsection{Why Does \tool Work?}
We identify 
the following two advantages of \tool, which can explain its effectiveness in code vulnerability detection. We also show an example of the correct prediction by \tool.

\textbf{(1) The ability to capture global information of code structure graph.} The proposed graph simplification method and kernel-scaled representation module help \tool learn the global information of code structure graphs.
More specifically, the proposed graph simplification methods reduce the node distance and code structure graph size, which can help \tool more effectively extract global information from code graphs. As shown in Figure~\ref{distance}, compared with the original code structure graph, the average node distance and maximum node distance on three datasets drop by 41.65\% and 39.68\%, respectively. 
Figure~\ref{cs_result} also shows that the numbers of nodes and edges are reduced after simplification. 
The proposed Kernel-scaled representation (KSR) module adopts two convolution kernels with different scales: a large kernel and a small kernel. The large kernel can focus on the relations between distant nodes, which benefits the performance of \tool. As shown in Table~\ref{table_rq3}, KSR module brings  obvious performance improvement to vulnerability detection, \eg 32.86\% in terms of F1 score on three datasets on average.  

\textbf{(2) The ability to handle graphs with multiple edge types.} 
We design the EA-GCN 
module for heterogeneous multi-relational code graphs. It can produce the edge-enhanced node representation matrix, which incorporates
the edge information in the simplified code structure graph. As shown in Table \ref{table_rq5}, the EA-GCN 
module has significant advantages over all the other GNN models, with an average improvement of 2.39\% and 7.40\% in accuracy and F1 score, respectively.


To understand whether \tool
pays attention to the vulnerable statements in the code, we illustrate the heatmap of attention for one example of CWE-77~\cite{CWE-77} in Figure~\ref{vul_example}. 
Statements with higher attention weights are shaded in red.
Following Reveal~\cite{reveal}, we obtain the node importance by using an activation function. The activation values of each statement and its corresponding sub-AST nodes are summed as the statement's `attention' weight. Lines 8-9 are vulnerable statements and the heatmap shows that \tool focuses more on Line 8. 
Since the code in Lines 4-9 has strong semantic dependencies with Lines 8 and 9, these lines also receive relatively higher
attention, which
enables \tool to make the accurate detection (\ie classifying the code as vulnerable).

In this paper, the proposed techniques benefit the graphs that contain multiple types of edges. We will investigate other types of code graphs and other tasks in our future work.


\subsection{Threats and Limitations}
The first threat to validity is about the limited number of experimental datasets. {We evaluate \tool on three datasets: FFMPeg+Qemu, Reveal, and Fan \et. The main reason is that these three datasets are commonly used by previous related work \cite{devign, reveal,IVDETECT} on vulnerability detection.} 
{Although these datasets cover many different types of open source projects, we will experiment with more datasets to further evaluate the effectiveness of \tool in our future work. }

The second threat to validity is that our proposed graph simplification is specific
to C/C++ languages.  
We only conduct experiments on C/C++ datasets, and not on datasets of other programming languages such as Java and C\#. In the future, we will select datasets in more programming languages to further evaluate our approach.

The third threat to validity is the implementation of baselines. Since Devign~\cite{devign} does not publish their implementation and hyper-parameters, we
reproduce Devign based on the Reveal's~\cite{reveal} implementation.
We try our best to tune its parameters as much as we can to achieve the similar experimental results reported in their paper. 



\section{Conclusion}
\label{sec:conclusion}
This paper proposes \tool, a novel vulnerability detection framework with graph simplification and enhanced graph representation learning.
\tool can shrink the node sizes of the code structure graphs to reduce the distances between nodes. 
It incorporates edge types to enhance the representation of local nodes to cope with more heterogeneous multi-relations in node representations. It also can capture the knowledge of graphs by capturing the relations between distant graph nodes. Compared with state-of-the-art deep learning-based methods, \tool significantly improves the vulnerability detection performance on all datasets by 7.64\%-199.81\% with respect to the F1 score. 


\textbf{Data availability}: Our source code as well as experimental data are available at: \textit{{\http}}.






\bibliographystyle{IEEEtran} 

\bibliography{IEEEabrv, Citation}


\end{document}